\documentclass[11pt]{article}
\usepackage{jheppub}
\usepackage{axodraw}
\usepackage{psfrag} 
\usepackage{bbm} 
\allowdisplaybreaks 

\newcommand{\la}[1]{\label{#1}}
\newcommand{\fig}{Fig.~}
\newcommand{\eq}{Eq.~}
\newcommand{\se}{Sec.~}
\newcommand{\eqs}{Eqs.~}
\newcommand{\nr}[1]{(\ref{#1})}
\renewcommand{\vec}[1]{{\bf #1}}
\renewcommand{\(}{\left(}
\renewcommand{\)}{\right)}
\newcommand{\lb}{\left\{}
\newcommand{\rb}{\right\}}
\newcommand{\lk}{\left[}
\newcommand{\rk}{\right]}
\newcommand{\e}{\epsilon}
\newcommand{\sumint}[1]{\hbox{$\sum$}\!\!\!\!\!\!\!\int_{#1}}
\newcommand{\dA}{d_{\mathrm{A}}}
\newcommand{\CA}{C_{\mathrm{A}}}
\newcommand{\CF}{C_{\mathrm{F}}}
\newcommand{\Nf}{N_{\mathrm{f}}}
\newcommand{\Nc}{N_{\mathrm{c}}}
\newcommand{\II}{I}
\newcommand{\Ib}[2]{\II_{#1}^{#2}}
\newcommand{\If}[2]{\hat \II_{#1}^{#2}}
\newcommand{\gM}{g_{\mathrm{M}}}

\newcommand{\mE}{m_{\mathrm{E}}}
\newcommand{\lE}{\lambda_{\mathrm{E}}}
\def\Iba{B_1}
\def\Ibb{B_2}
\def\Ibc{B_3}
\def\Ibd{B_4}
\def\Ibe{B_5}
\def\Ifa{F_1}
\def\Ifb{F_2}
\def\Ifc{F_3}
\def\Ifd{F_4}
\def\Ife{F_5}
\def\Iff{F_6}
\def\Ifg{F_7}
\def\Ifh{F_8}
\def\Ifi{F_9}
\def\Ifj{F_{10}}
\newcommand{\gR}{g_{\rm R}}
\newcommand{\zA}{Z_1}
\newcommand{\zC}{Z_3}
\newcommand{\rmii}[1]{{\mbox{\tiny\rm{#1}}}}
\newcommand{\gaE}{\gamma_{\rmii E}}
\newcommand{\fra}[2]{\mbox{\small$\frac{#1}{#2}$}}

\newcommand{\pic}[1]{\;\parbox[c]{30pt}{\begin{picture}(30,30)(0,0)
\SetWidth{1.0}\SetScale{1.0} #1 \end{picture}}\;}
\newcommand{\picb}[1]{\;\parbox[c]{45pt}{\begin{picture}(45,30)(0,0)
\SetWidth{1.0}\SetScale{1.0} #1 \end{picture}}\;}
\newcommand{\picc}[1]{\;\parbox[c]{60pt}{\begin{picture}(60,30)(0,0)
\SetWidth{1.0}\SetScale{1.0} #1 \end{picture}}\;}
\def\Lwidth{1}

\def\Agl(#1,#2)(#3,#4,#5){\PhotonArc(#1,#2)(#3,#4,#5){\Lwidth}
{6.283 #3 mul 360 div #4 #5 sub #4 #5 sub mul sqrt mul Ldensity mul}}
\def\Lgl(#1,#2)(#3,#4){\Photon(#1,#2)(#3,#4){\Lwidth}
{#1 #3 sub #1 #3 sub mul #2 #4 sub #2 #4 sub mul add sqrt Ldensity mul}}
\def\Ahg(#1,#2)(#3,#4,#5){\DashArrowArcn(#1,#2)(#3,#5,#4){1}}
\def\Lhg(#1,#2)(#3,#4){\DashArrowLine(#3,#4)(#1,#2){1}}
\def\Auq(#1,#2)(#3,#4,#5){\ArrowArcn(#1,#2)(#3,#5,#4)}
\def\Luq(#1,#2)(#3,#4){\ArrowLine(#3,#4)(#1,#2)}
\def\TopoVR(#1){\pic{#1(15,15)(15,-90,270)}}
\def\ToptVS(#1,#2,#3){\pic{#1(15,15)(15,0,180) #2(15,15)(15,180,360)%
 #3(30,15)(0,15)}}
\def\ToptVE(#1,#2){\picc{#1(15,15)(15,0,360) #2(45,15)(15,-180,180)}}
\def\ToprVM(#1,#2,#3,#4,#5,#6){\pic{#3(15,15)(15,-30,90) #1(15,15)(15,90,210)%
 #2(15,15)(15,210,330) #5(15,15)(2,7.5) #6(15,30)(15,15) #4(28,7.5)(15,15)}}
\def\ToprVV(#1,#2,#3,#4,#5){\!\!\picb{#2(26.25,15)(15,256,76)%
 #3(30,30)(15,30) #1(18.75,15)(15,104,284) #4(15,30)(22.5,0)%
 #5(30,30)(22.5,0)}\!\!}
\def\ToprVB(#1,#2,#3,#4){\picb{#1(30,15)(15,-120,120) #2(30,15)(15,120,240)%
 #3(15,15)(15,60,300) #4(15,15)(15,-60,60)}}
\def\ToprVBB(#1,#2,#3,#4,#5,#6){\picb{#1(30,15)(15,-90,90)%
 #2(15,30)(30,30) #3(15,15)(15,90,270) #4(30,0)(15,0)%
 #5(30,30)(30,0) #6(15,0)(15,30)}}
\def\ToprVBT(#1,#2,#3,#4,#5){\picc{#3(15,15)(15,0,360) #5(45,30)(45,0)
 #1(45,15)(15,-90,90) #2(45,15)(15,90,180) #4(45,15)(15,180,270)}}
\def\ToprVTT(#1,#2,#3,#4){\picc{#2(30,15)(15,0,180) #4(30,15)(15,180,360)%
 #1(52.5,15)(7.5,-180,180) #3(7.5,15)(7.5,0,360)}}

\title{IBP methods at finite temperature}

\preprint{BI-TP 2012/28}

\author{Madoka Nishimura}
\author{and York Schr\"oder}

\affiliation{Tohoku University of Community Service and Science, 
  Sakata 998-8580, Japan}
\affiliation{Faculty of Physics, University of Bielefeld, 
  33501 Bielefeld, Germany}

\emailAdd{madoka.nishimura@koeki-u.ac.jp}
\emailAdd{yorks@physik.uni-bielefeld.de}

\abstract{We demonstrate the applicability of integration-by-parts (IBP)
identities in finite-temperature field theory. As a concrete example,
we perform 3-loop computations for the thermodynamic pressure of QCD in 
general covariant gauges, and confirm earlier Feynman-gauge results.}


\begin{document}
\maketitle
\flushbottom

%
\section{Introduction}
\la{se:intro}

Modern perturbative computations typically involve large numbers
of Feynman diagrams and/or Feynman integrals,
whose successful evaluation is greatly facilitated by the use of 
computer algebra.
Indeed, in the field of multi-loop calculations (in zero-temperature
field theories) the algorithmic formulation is in a quite mature
state, as witnessed by numerous higher-order results, such as e.g. 
5-loop results for the QED beta function \cite{Baikov:2012zm}, 
complete 3-loop and parts of 4-loop terms in the electron
anomalous magnetic moment \cite{Laporta:1996mq},
or 4-loop contributions to electroweak precision observables 
\cite{Schroder:2005db}, just to name a few.

Such calculations are typically performed in a sequence of steps:
(a) generation of the complete set of diagrams and counter-terms
contributing to the observable under study; 
(b) application of Feynman rules, necessary projectors and traces,
performing Lorentz algebra and scalarization;
(c) mapping onto a set of integral templates;
(d) reduction to a few (``master'') integrals;
(e) expansion in epsilon.

For each of these distinct steps, a variety of systematic algorithmic 
methods and tools can be applied, the most commonly used ones being:
(a) graph theory (efficiently coded in the package 
QGRAF \cite{Nogueira:1991ex});
(b,c) computer-algebra systems 
(such as e.g. FORM \cite{Vermaseren:2000nd}, 
Ginac \cite{Bauer:2000cp}, 
Mathematica \cite{mma});
(d) integration-by-parts methods (pioneered by Chetyrkin and 
Tkachov \cite{Chetyrkin:1981qh},
formalized by Laporta in \cite{Laporta:2001dd} 
and implemented e.g. in the public
packages Air \cite{Anastasiou:2004vj}, 
FIRE \cite{Smirnov:2008iw}, 
Reduze \cite{Studerus:2009ye});
(e) difference equations \cite{Laporta:2001dd},
differential equations (see, e.g. \cite{Laporta:2003xa}),
harmonic polylogarithms and -sums \cite{Remiddi:1999ew}
(where the expansion can be automatized for simple 0-scale 
problems \cite{Moch:2001zr},
but is non-trivial for multiple-scale problems),
graph polynomials \cite{Bogner:2010kv},
sector decomposition \cite{Heinrich:2008si}.

For problems in finite temperature field theories, however,
a similar computer-algebraic approach to higher-order calculations
is much less developed, although a limited number of results up to 
the three-loop \cite{Arnold:1994ps,Braaten:1995jr,Andersen:2011sf} 
and even the four-loop level \cite{Gynther:2007bw}
do exist. Whenever gauge theories are treated, owing to the 
structure of gauge-field propagators and -vertices, most
authors choose to work in a fixed (typically Feynman) gauge,
in order to reduce the complexity of the calculation.

While most of these works were performed in a more traditional
way, there is no major obstacle in carrying over the systematics
of the modern computer-algebraic
developments mentioned above to finite temperature calculations,
which as a consequence abolishes the need to work in fixed gauges.
In fact, for the steps (a) to (d) -- i.e. diagram generation / algebraic
simplifications / mapping / reduction --
the methods from zero-temperature field theory can be generalized
directly, with a few minor modifications.
It is only step (e) -- i.e. the epsilon expansion of master 
integrals -- that resists full automation, owing to the fact that
the available techniques for sum-integrals which are needed in the thermal 
setting are much more limited than those for pure continuum integrals
(for which at least numerical methods are guaranteed to work).
For first attempts in treating whole classes (as opposed to treating
them one-by-one as has been the state of the art 
previously \cite{Arnold:1994ps,Gynther:2007bw,Schroder:2008ex}) 
of non-trivial sum-integrals, see \cite{Moeller:2012da}.

In this paper, we want to demonstrate the utility of computer-algebra methods
for thermal field theories, which mainly concerns the integral 
reduction step (d). 
Concentrating on the concrete example of the 3-loop free energy 
of hot QCD, we briefly introduce the corresponding observable
in \se\ref{se:setup} and then explain the specifics of the (IBP) 
reduction method generalized to finite temperature 
in \se\ref{se:reduction}.
\se\ref{se:result} contains the result of our diagrammatic calculation,
confirming the known 3-loop result \cite{Braaten:1995jr},
however in general covariant gauges.
In \se\ref{se:conclusion} we conclude. 
The Appendix lists some useful sum-integrals that are needed
for the final result.

%
\section{Thermodynamic observables}
\la{se:setup}

QCD equilibrium properties, such as its free energy density $F$, 
are encoded in the logarithm of the partition function
\begin{align}
\la{eq:F}
&F = -\frac{T}{V}\,\ln \int\!\!{\cal D}[A_\mu,\bar\psi,\psi]\,
\exp\(-\int_0^{1/T}\!\!{\rm d}\tau\int\!\!{\rm d}^dx\,
{\cal L}_{\rm QCD}\) \;.
\end{align}
Here, $V=\int\!\!{\rm d}^dx$ denotes the spatial volume occupied by the 
system ($d=3\!-\!2\e$), 
and we work in the imaginary time formalism, defined on a ($d$+1)-dimensional
Euclidean space with a compact temporal coordinate $\tau$
with period $1/T$ set by the temperature. 
Bosonic/fermionic (gluon/quark) fields are periodic/antiperiodic
functions of $\tau$. The (Euclidean) QCD Lagrangian reads 
\begin{align}
\la{eq:L_QCD}
&{\cal L}_{\rm QCD} = \frac14\,F_{\mu\nu}^a F_{\mu\nu}^a
+\sum_{i=1}^{\Nf}\bar\psi_i \(\gamma_\mu D_\mu + m_i + i\mu_i\)\psi \;,
\end{align}
with field strength tensor 
$F_{\mu\nu}^a=\partial_\mu A_\nu^a-\partial_\nu A_\mu^a+gf^{abc}A_\mu^b A_\nu^c$
and covariant derivative $D_\mu=\partial_\mu+igA_\mu^a T^a$.
We will set quark masses $m_i$ and chemical potentials $\mu_i$
to zero here, and work with the gauge group SU($\Nc$),
where $\CA=\Nc$, ${\rm Tr}(T^aT^b)=\frac12\,\delta^{ab}$ and
$T^aT^a=C_F\mathbbm{1}=\frac{\Nc^2-1}{2\Nc}\,\mathbbm{1}$.

It turns out that the weak-coupling expansion of \eq\nr{eq:F}
is nonanalytic in the strong coupling constant $\alpha_s=g^2/4\pi$.
The physical reason is that, due to multiple interactions in 
the thermal medium, screening masses are dynamically generated 
for all massless particles, resulting in a multi-scale system.
In fact, at high temperatures, asymptotic freedom guarantees a small 
gauge coupling $g$. In this regime, QCD develops a hierarchy of 
three momentum scales $\pi T\gg gT \gg g^2T/\pi$,
whose effect can be most transparently accounted for in an 
effective theory setup \cite{Braaten:1995cm}.
Systematically integrating out the largest (``hard'') scale $\pi T$
and the second-largest (``soft'') scale $gT$ in turn, 
one obtains a dimensionally reduced effective theory 
\cite{Ginsparg:1980ef}
for the smallest (``ultrasoft'') scale, which has
been dubbed magnetostatic QCD (MQCD) \cite{Braaten:1995jr}.
The leading term of the MQCD action turns out to be a 3-dimensional
pure Yang-Mills theory, which is confining and therefore has
to be treated with suitable non-perturbative methods \cite{Hietanen:2004ew}.

Utilizing the effective theory setup, the QCD pressure 
can be expressed as \cite{Braaten:1995jr}
\begin{align}
\la{eq:p_hsu}
p_{\rm QCD}(T)
&\equiv -\lim_{V\rightarrow\infty}F
\;=\;p_{\rm hard}(T)+p_{\rm soft}(T)+p_{\rm ultrasoft}(T) \;,
\end{align}
where $p_{\rm hard}$ and $p_{\rm soft}$ are perturbatively computable
matching coefficients which account for contributions from hard
and soft momentum scales, respectively.
Defining these perturbative matching coefficients in the $\overline{\rm MS}$
scheme, the remaining contribution
\begin{align}
\la{eq:p_u}
p_{\rm ultrasoft}(T)\equiv\lb\lim_{V\rightarrow\infty}\frac{T}{V}\,
\ln\int\!\!{\cal D}[A_i]\,\exp\(-\int\!\!{\rm d}^dx\,{\cal L}_{\rm MQCD}\)
\rb^{\overline{\rm MS}}
\end{align}
entails the effective Lagrangian 
${\cal L}_{\rm MQCD}=\frac14\,F_{ij}^aF_{ij}^a+\dots$,
where $F_{ij}^a=\partial_i A_j^a-\partial_j A_i^a+\gM f^{abc}A_i^b A_j^c$
contains the 3-dimensional gauge coupling $\gM$. 
In contrast to the dimensionless 4-dimensional gauge coupling $g$,
the dimensionality of $\gM^2$ is GeV,
such that for dimensional reasons $p_{\rm ultrasoft}\sim T\,\gM^6$.
On the other hand, perturbative matching yields 
$\gM^2=g^2 T\(1+{\cal O}(g)\)$,
so $p_{\rm ultrasoft}$ plays a role in $p_{\rm QCD}$ starting at 
${\cal O}(g^6)$ only, which is beyond the precision needed for
our investigation (see, however \cite{Hietanen:2004ew}).

The two matching coefficients in \eq\nr{eq:p_hsu}
are defined in analogy to \eq\nr{eq:p_u}, but with actions 
$\int_0^{1/T}\!\!{\rm d}\tau\int\!\!{\rm d}^dx\,{\cal L}_{\rm QCD}$ 
and 
$\int\!\!{\rm d}^dx\,{\cal L}_{\rm EQCD}$ 
for $p_{\rm hard}$ and $p_{\rm soft}$, respectively.
The latter depends on the so-called 3d electrostatic QCD (EQCD) Lagrangian
containing a massless gauge field $A_i$ and
a massive adjoint scalar $A_0$,
and whose structure results from integrating out the hard scales from QCD, 
yielding the leading terms
${\cal L}_{\rm EQCD}=\frac14\, F_{ij}^aF_{ij}^a+{\rm Tr}[D_iA_0][D_iA_0]+
\mE^2{\rm Tr}A_0^2+\lE^{(1)}\({\rm Tr}A_0^2\)^2
+\lE^{(2)}{\rm Tr}A_0^4+\dots$.
For a review on the status of the different contributions we refer to
\cite{Schroder:2006vz}. 

It turns out that at present, the bottleneck is the evaluation
of higher orders in $p_{\rm hard}(T)$, which can be obtained by 
adding all vacuum diagrams in 4-dimensional thermal QCD, evaluated
in the ``naive'' perturbative sense, i.e. by regularizing ultraviolet
as well as infrared divergences in dimensionally 
(it is the effective theory setup \eq\nr{eq:p_hsu} that properly
accounts for infrared effects \cite{Linde:1980ts} that need to be resummed).
The need to ultimately perform 4-loop computations within this 
setting is our main motivation to proceed with computer-algebraic
methods as far as possible. As mentioned in the introduction, 
this concerns mainly the task of integral reduction, which we
will now discuss, and then apply to the problem of 3-loop
corrections to $p_{\rm hard}(T)$, enabling a comparison with known
results from the literature.
Going a step beyond the Feynman-gauge treatment of \cite{Braaten:1995jr}, 
we work in covariant gauges with gluon propagator
\begin{align}
\la{eq:propag}
D_{\mu\nu}^{ab}(P) = 
\delta^{ab}\lk\frac{\delta_{\mu\nu}}{P^2} - 
\xi\frac{P_{\mu}P_{\nu}}{(P^2)^2}\rk\;,
\end{align}
explicitly keeping the gauge parameter $\xi$
(note that in our convention, $\xi=0(1)$ corresponds to
Feynman (Landau) gauge, respectively), and 
demonstrating its cancellation in the sum of diagrams.

%
\section{Integral reduction}
\la{se:reduction}

For the sake of concreteness, and to 
avoid too generic notation and proliferation of indices,
let us take the problem of three-loop sum-integral reduction
as an example here.
While it should be understood that the methods introduced below
are independent of this choice, let us note that this is in
fact the first non-trivial loop order in the case of sum-integrals
(cf. Appendix~\ref{app:12loop}), and also precisely the level 
of the computation displayed in \se\ref{se:result}.

In general, 3-loop vacuum-type sum-integrals 
can be written in the form
\begin{align}
\la{eq:sc3l}
I_{abcdef;\,c_pc_qc_r}^{\alpha\beta\gamma} \equiv 
\sumint{PQR}
\frac{(P_0)^{\alpha}\,(Q_0)^{\beta}\,(R_0)^{\gamma}}
{[P^{2}]^{a}\,[Q^{2}]^{b}\,[R^{2}]^{c}\,
[(P-Q)^{2}]^{d}\,[(P-R)^{2}]^{e}\,[(Q-R)^{2}]^{f}}\,,
\end{align}
where $P^2 = (P_0)^2 + \vec{p}^{2} = ([2n_p+c_p]\pi T)^2+\vec{p}^{2}$ 
are bosonic (fermionic) loop momenta for $c_i=0$ ($1$),
and where the indices $a\dots f\in\mathbbm{Z}$ and 
$\alpha\dots \gamma\in\mathbbm{N}_0$.
The sum-integral symbol
in \eq\nr{eq:sc3l} is a shorthand for
\begin{align}
\sumint{P} \equiv
\mu^{2\e}T\sum_{P_{0}}\int\frac{\mathrm{d}^{d}\vec p}{(2\pi)^{d}}\,,
\end{align}
where $\mu$ is the minimal subtraction ($\text{MS}$) 
scheme scale parameter, we take $d=3-2\e$, 
and the sum is over all integers $n_p\in\mathbbm{Z}$.
Hence, the set of indices of the sum-integral $I$ 
enumerates all possible structures that can occur in a particular
(3-loop) computation of finite-temperature Feynman integrals 
without further external momentum scales (a generalization to
$n$\/-point functions or to 
internal masses and/or chemical potentials is straightforward, 
but let us focus on the problem at hand here).

Generic integration-by-parts (IBP) relations then provide
linear relations between the set of sum-integrals $I$
of \eq\nr{eq:sc3l}, using that the integral of a total derivative
(here with respect to the spatial loop momenta only)
vanishes in dimensional regularization,
\begin{align}
\la{eq:IBP}
0 &=
\sumint{P\dots} \partial_{p_i}\,f_i(p_i,P_0,\dots)\,g(P,\dots)\,,
\end{align}
with arbitrary function $f_i$ (and where $g$ denotes an integrand of the 
type \eq\nr{eq:sc3l}).
The linear relations are obtained by choosing the function $f_i$,
working out the derivatives, and re-expressing the result in terms
of the generic form \eq\nr{eq:sc3l}.
From this point on, it is clear that the well-established 
(zero-temperature) algorithms that systematically solve such systems 
of linear IBP relations can be taken over. In practice, providing
a unique ordering relation among the sum-integrals $I$, we use two 
Laporta-type \cite{Laporta:2001dd} algorithms (one programmed in 
FORM \cite{Vermaseren:2000nd}, 
as well as one in Ruby \cite{ruby}; the latter code is used as 
a cross-check on the former, setting the dimension $d$ to a numerical value,
utilizing the speed of numerical Gaussian elimination, and avoiding
polynomial algebra).

Furthermore, to reduce the number of relations, it is useful
to exploit symmetries among the $I$, which can be generated
by linear shifts of loop-momenta.
For example,
due to the symmetry of the generic sum-integral in \eq\nr{eq:sc3l}
it is sufficient to consider, from all $2^3=8$ possibilities,
the three cases $(c_p,c_q,c_r)=\{(000),(100),(110)\}$.
This mapping can be automated by systematically 
treating linear relations originating from momentum shifts on the 
same footing as the IBP relations.
A typical relation originating from such shifts is
$I_{000111;001}^{000}=I_{111000;110}^{000}$. 
In general, however, for non-zero coefficients $\alpha,\beta,\gamma$
and/or negative values among the indices $a-f$ 
(i.e. scalar products in the numerator), these relations
have more than one term on the right-hand side.

One could suspect that there exist additional IBP-type relations that take
into account the structure of the Matsubara sums, which \eq\nr{eq:IBP}
has not sampled yet. This is not the
case, however, as we will show here. 
In fact, acting with the operator $T\partial_T$ on
a generic $L$\/-loop sum-integral (assuming massless propagators,
which however can be of bosonic or fermionic type),
one can either use the fact that, 
due to the absence of any other dimensional scale,
the dimension of the sum-integral
is carried by the scale $T$ only, or apply the derivative to the 
explicit (in front of each sum) and implicit 
(in the $P_0=(2n+c)\pi T$ of the integrand) 
occurrences of $T$ directly
\begin{align}
\la{eq:xtraIBP}
0&=
\bigg\{T\partial_T-T\partial_T\bigg\}\;
T\sum_{P_{10}}\dots T\sum_{P_{L0}}
\int_{p_1}\dots \int_{p_L}
g(P_{1},\dots,P_{L})\nonumber\\
&=
\sumint{P_1\dots P_L} 
\bigg\{\mbox{Dim}_{\rm Int}-L-\sum_{i=1}^L P_{i0}\partial_{P_{i0}}\bigg\}\;
g(P_{1},\dots,P_{L})\,.
\end{align}
It turns out that 
\eq\nr{eq:xtraIBP} carries the same information as the sum of the
``diagonal'' IBP relations
\begin{align}
0&=\sumint{P_1\dots P_L}\bigg\{\partial_{\vec p_1}\!\!\cdot\!\vec p_1
+\partial_{\vec p_2}\!\!\cdot\!\vec p_2
+\dots+\partial_{\vec p_L}\!\!\cdot\!\vec p_L\bigg\}\;g(P_1,\dots,P_L) \;,
\end{align}
and therefore does not need to be considered.
We have checked this explicitly up to four loops, for vacuum sum-integrals. 

One can however use an additional set of relations that derive
from the sum-part of the sum-integrals. 
These additional relations 
essentially mix bosonic with fermionic sum-integrals,
and are based on
scaling arguments, such as used e.g. in Appendix B of \cite{Arnold:1992rz}
(see also \cite{Coriano:1994re}).
One first re-scales the spatial integration momenta 
of a given bosonic integral as $p_i\rightarrow 2p_i$
and then partitions the Matsubara sums as
\begin{align}
\la{eq:PC}
\sum_{n\in\mathbbm{Z}}=\sum_{\rm even}+\sum_{\rm odd} 
\;.
\end{align}
In practice, this provides a few linear relations
among different bosonic and fermionic master sum-integrals,
which remain after systematic use of the IBP relations \eq\nr{eq:IBP}.
The simplest example is the relation between 1-loop tadpoles, 
as shown in \eq\nr{eq:Ifmn}.
For the 3-loop calculation presented in \se\ref{se:result} below,
we obtain (letting $I_{110011;abc}^{000}\equiv I_{abc}$ for brevity)
\begin{align}
\la{eq:PC3}
I_{000} &=\sumint{PQR}\frac1{P^2\,Q^2\,(P-R)^2\,(Q-R)^2}
\nonumber\\
&=T^3
\sum_{n_p\in\mathbbm{Z}}
\sum_{n_q\in\mathbbm{Z}}
\sum_{n_r\in\mathbbm{Z}}
\int_{\vec p\vec q\vec r}
\frac1{[(2n_p\pi T)^2+\vec p^2]\,[(2n_q\pi T)^2+\vec q^2]
\dots}\nonumber\\
&=2^{3d-8}\;T^3
\sum_{n_p\in\mathbbm{Z}}
\sum_{n_q\in\mathbbm{Z}}
\sum_{n_r\in\mathbbm{Z}}
\int_{\vec p\vec q\vec r}
\frac1{[(n_p\pi T)^2+\vec p^2]\,[(n_q\pi T)^2+\vec q^2]
\dots}\nonumber\\
&=2^{3d-8}\;\Big(
I_{000}+I_{001}+I_{010}+I_{100}+I_{110}+I_{101}+I_{011}+I_{111}
\Big)\nonumber\\
&=2^{3d-8}\;\Big(
I_{000}+6\,I_{001}+I_{110}
\Big)\;,
\end{align}
where in the third line we have scaled the spatial momenta,
in the fourth line considered all cases of \eq\nr{eq:PC} (cubed),
and finally exploited symmetries of this simple basketball-type
sum-integral. Altogether, we therefore have the linear relation
(anticipating the notation of \eqs\nr{eq:mastersIb}ff for master integrals)
\begin{align}
\la{eq:PCrel}
0&=(1-2^{8-3d})\,I_{000}+6\,I_{001}+I_{110} \;=\;
(1-2^{8-3d})\,\Ibb+6\,\Iff+\Ifg
\end{align}
between one bosonic and two fermionic integrals,
which will allow us to reduce the basis of master integrals by one.
We feed these types of linear relations into our IBP system as well.

%
\section{Results of diagrammatic calculation}
\la{se:result}

Let us now evaluate the coefficient $p_{\rm hard}(T)$, 
as defined in \se\ref{se:intro}, to three-loop order.
In doing so, we work in covariant gauges 
(cf \eq\nr{eq:propag}), aiming at proving gauge parameter 
independence as well as confirming the corresponding 
Feynman-gauge result of \cite{Braaten:1995jr}.
Working in dimensional regularization with $d=3-2\e$, 
let us rewrite bare quantities as 
\begin{align}
p_{\rm h}(T)\equiv\mu^{2\e}p_{\rm hard}^{\rm bare}(T)\;,\qquad
g^2\equiv\mu^{-2\e}g_{\rm bare}^2\;.
\end{align}

\begin{figure}
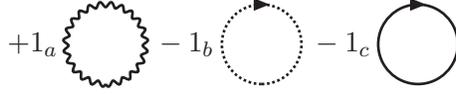

\begin{eqnarray*}
&&
 +{1}_{a}\TopoVR(\Agl)
 -{1}_{b}\TopoVR(\Ahg)
 -{1}_{c}\TopoVR(\Auq)
\end{eqnarray*}
\caption{``One-loop'' contributions to the QCD pressure. 
Wiggly/dotted/solid lines denote gluons/ghosts/quarks, respectively.
The alphabetical index on the prefactor labels the diagram.}
\la{fig:1loop}
\end{figure}

The leading (``one-loop'' -- or, more precisely, the logarithms of
Gaussian path integrals over the quadratic parts of the action, 
given by the sum over logarithms of momentum-space eigenvalues
of the corresponding matrix kernels) terms of \fig\nr{fig:1loop}
can be expressed in terms of basic logarithmic integrals,
which in turn can be related to the more conventional one-loop
sum-integrals over propagator structures $\Ib{m}{n}$
given in \eq\nr{eq:Imn}. For the bosonic case,
\begin{align}
\la{eq:logP}
I \equiv \sumint{P}\ln(P^2) \quad
\Rightarrow\quad T^2\partial_{T^2} I = \frac{d+1}2\,I = 
\frac12\,I+\sumint{P}\frac{P_0^2}{P^2}\quad
\Leftrightarrow\quad I=\frac2d\,\sumint{P}\frac{P_0^2}{P^2}=\frac2d\,\Ib{1}{2}
\;.
\end{align}
Here, we noticed that $I$ depends on the scale $T$ only, such that
on the one hand its derivative gives the overall dimension,
while on the other hand the derivative can be applied directly
(hitting the explicit $T$ in the sum-integral as well as the implicit 
one in $P^2$).
In complete analogy, one gets for the fermionic case
\begin{align}
\hat I \equiv \sumint{P_f}\ln(P^2) \;=\; \frac2d\,\If{1}{2}
\;=\; \frac2d\(2^{-d}-1\)\Ib{1}{2}\;,
\end{align}
such that the individual diagrams of \fig\nr{fig:1loop}
contribute as
\begin{alignat}{2}
p_{\rm h}^{[a]}&=-\frac{d+1}{d}\,\dA\Ib{1}{2}&&\;\stackrel{d=3}=\; 
  T^4\frac{2\pi^2}{45}\,\dA\;,\\
p_{\rm h}^{[b]}&=\frac{2}{d}\,\dA\Ib{1}{2}   &&\;\stackrel{d=3}=\;
 -T^4\frac{\pi^2}{45}\,\dA\;,\\
p_{\rm h}^{[c]}&=\frac{4}{d}\(2^{-d}\!-\!1\)\Nf\CA\Ib{1}{2}&&\;
  \stackrel{d=3}=\; 
  T^4\frac{7\pi^2}{180}\,\Nf\CA\;,
\end{alignat}
whose sum gives the well-known (QCD version of the) Stefan-Boltzmann law.
One can see clearly the effect of the ghosts here, which cancel half
of the result of the pure gluonic contribution. 

\begin{figure}
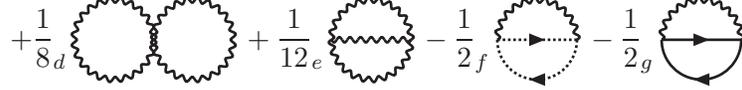

\begin{eqnarray*}
&&
 +{1\over8}_{d}\ToptVE(\Agl,\Agl)
 +{1\over12}_{e}\ToptVS(\Agl,\Agl,\Lgl)
 -{1\over2}_{f}\ToptVS(\Agl,\Ahg,\Lhg)
 -{1\over2}_{g}\ToptVS(\Agl,\Auq,\Luq)
\end{eqnarray*}
\caption{Two-loop contributions to the pressure. 
Notation as in \fig\nr{fig:1loop}}
\la{fig:2loop}
\end{figure}

After reduction of the 2-loop diagrams of \fig\nr{fig:2loop}, 
we obtain $d$-dimensional bare results, expressed in terms of bosonic
as well as fermionic 1-loop tadpoles (as has been pointed out e.g. in 
\cite{Braaten:1995jr}, all 2-loop sum-integrals factor into products 
of two 1-loop cases; this observation we reproduce
via IBP, for all possible 2-loop vacuum sum-integrals, also
of different dimensionality needed for other computations; 
see also Appendix \ref{app:12loop})
\begin{align}
   p_{\rm h}^{[d]}&=
      g^2\dA\CA\left(
         \frac{d(d-3)}{16}\,\xi^2
         +\frac{d}{2}\,\xi
         -\frac{d(d+1)}{4}
         \right)\Ib{1}{0}\Ib{1}{0}\;,\\
   p_{\rm h}^{[e]}&=
      g^2\dA\CA\left(
         -\frac{d(d-3)}{16}\,\xi^2
         -\frac{4d+1}{8}\,\xi
         +\frac{3d}{4}
         \right)\Ib{1}{0}\Ib{1}{0}\;,\\
   p_{\rm h}^{[f]}&=
      g^2\dA\CA\left(
         \frac{1}{8}\,\xi
         -\frac{1}{4}
         \right)\Ib{1}{0}\Ib{1}{0}\;,\\
   p_{\rm h}^{[g]}&=
      g^2\dA\Nf\frac{d-1}{2}\left(
         2\Ib{1}{0}
         -\If{1}{0}
         \right)\If{1}{0}
\;=\; g^2\dA\Nf\frac{d-1}{2}\(2^{4-d}-4^{2-d}-3\)\Ib{1}{0}\Ib{1}{0}
\;.
\end{align}

\begin{figure}
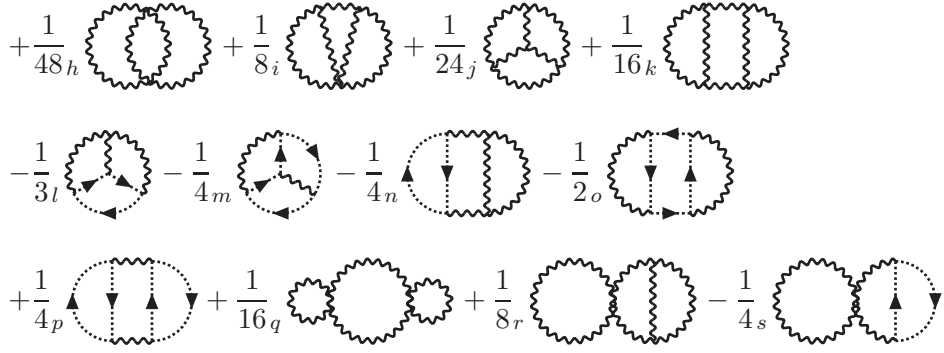

\begin{eqnarray*}
&&
 +{1\over48}_{h}\ToprVB(\Agl,\Agl,\Agl,\Agl)
 +{1\over8}_{i}\ToprVV(\Agl,\Agl,\Lgl,\Lgl,\Lgl)
 +{1\over24}_{j}\ToprVM(\Agl,\Agl,\Agl,\Lgl,\Lgl,\Lgl)
 +{1\over16}_{k}\ToprVBB(\Agl,\Lgl,\Agl,\Lgl,\Lgl,\Lgl)
\\[5mm]&&
 -{1\over3}_{l}\ToprVM(\Agl,\Ahg,\Agl,\Lhg,\Lhg,\Lgl)
 -{1\over4}_{m}\ToprVM(\Agl,\Ahg,\Ahg,\Lgl,\Lhg,\Lhg)
 -{1\over4}_{n}\ToprVBB(\Agl,\Lgl,\Ahg,\Lgl,\Lgl,\Lhg)
 -{1\over2}_{o}\ToprVBB(\Agl,\Lhg,\Agl,\Lhg,\Lhg,\Lhg)
\\[5mm]&&
 +{1\over4}_{p}\ToprVBB(\Ahg,\Lgl,\Ahg,\Lgl,\Lhg,\Lhg)
 +{1\over16}_{q}\ToprVTT(\Agl,\Agl,\Agl,\Agl)
 +{1\over8}_{r}\ToprVBT(\Agl,\Agl,\Agl,\Agl,\Lgl)
 -{1\over4}_{s}\ToprVBT(\Ahg,\Agl,\Agl,\Agl,\Lhg)
\end{eqnarray*}
\caption{Three-loop diagrams contributing to $\CA^2$.}
\la{fig:3loop}
\end{figure}

Turning now to the three-loop case, the IBP reduction 
leaves us with a set of bosonic 
\begin{alignat}{2}
\la{eq:mastersIb}
\Iba&\equiv I_{112000;000}^{000}=\Ib{1}{0}\,\Ib{1}{0}\,\Ib{2}{0}\;,
&\Ibb&\equiv I_{110011;000}^{000}\;,\\
\Ibc&\equiv I_{21111-\!2;000}^{000}\;.
\qquad\Ibd\equiv I_{110012;000}^{200}\;,\qquad
&\Ibe&\equiv I_{11-\!2013;000}^{000}\;,
\end{alignat}
as well as fermionic master sum-integrals 
of the form \eq\nr{eq:sc3l}
(note however the linear relation \eq\nr{eq:PCrel} that can still be used
to eliminate one of $\Ibb,\Iff,\Ifg$)
\begin{alignat}{2}
\Ifa&\equiv I_{121000;001}^{000}=\Ib{1}{0}\,\Ib{2}{0}\,\If{1}{0}=\big(2^{2-d}-1\big)\Iba\;,
&\Iff&\equiv I_{110011;001}^{000}\;,\\
\Ifb&\equiv I_{112000;001}^{000}=\Ib{1}{0}\,\Ib{1}{0}\,\If{2}{0}=\big(2^{4-d}-1\big)\Iba\;,
&\Ifg&\equiv I_{110011;110}^{000}\;,\\
\Ifc&\equiv I_{112000;011}^{000}=\Ib{1}{0}\,\If{1}{0}\,\If{2}{0}=\big(2^{2-d}-1\big)\big(2^{4-d}-1\big)\Iba\;,
&\Ifh&\equiv I_{21111-\!2;001}^{000}\;,\\
\Ifd&\equiv I_{211000;011}^{000}=\Ib{2}{0}\,\If{1}{0}\,\If{1}{0}=\big(2^{2-d}-1\big)^2\Iba\;,
&\Ifi&\equiv I_{21111-\!2;011}^{000}\;,\\
\Ife&\equiv I_{112000;111}^{000}=\If{1}{0}\,\If{1}{0}\,\If{2}{0}=\big(2^{2-d}-1\big)^2\big(2^{4-d}-1\big)\Iba\;,\quad
&\Ifj&\equiv I_{11111-\!1;100}^{000}\;.
\end{alignat}
In terms of these masters, we obtain $d$-dimensional bare results
from the three-loop diagrams depicted in \fig\nr{fig:3loop}:
\begin{align}
\la{eq:hijk}
   p_{\rm h}^{[h+i+j+k]}&=
      g^4\dA\CA^2\bigg(
         \Iba\bigg(\frac{d(2+33 d-12 d^2+d^3)}{384}\,\xi^4
         +\frac{d(-6-13 d+3 d^2)}{32}\,\xi^3
+\nonumber\\&
         +\frac{2+41 d+56 d^2-8 d^3}{64}\,\xi^2
         -\frac{13+115 d+38 d^2-50 d^3+8 d^4}{8(d-3)(d-5)}\,\xi
         -\frac{3-14 d+4 d^2}{8}\bigg)
+\nonumber\\&
         +\Ibb\left(\frac{22-13 d+3 d^2}{768}\,\xi^2
         -\frac{11-19 d+6 d^2}{192(d-3)}\,\xi
         +\frac{5-10 d+4 d^2}{32}\right)
+\nonumber\\&
         +\Ibc\frac{(2d-1)^2}{16}
         -\Ibd\left(\frac{d-3}{32}\,\xi^2
         -\frac{1}{4}\,\xi\right)
         -\Ibe\left(\frac{1}{64}\,\xi^2
         -\frac{1}{8(d-3)}\,\xi\right)
         \bigg)\;,\\
\la{eq:l}
   p_{\rm h}^{[l]}&=
      g^4\dA\CA^2\bigg(
         \Iba\left(
         \frac{d-2}{16}\,\xi^2
         -\frac{d-2}{8(d-3)}\,\xi\right)
-\nonumber\\&
         -\Ibb\left(
         \frac{(d-1)(3d-10)}{384}\,\xi^2
         +\frac{28-17 d+3 d^2}{96(d-3)}\,\xi
         +\frac{1}{16}\right)
+\nonumber\\&
         +\Ibd\left(
         -\frac{d-3}{8}\,\xi^2
         +\frac{1}{4}\,\xi\right)
         +\Ibe\left(
         -\frac{1}{16}\,\xi^2
         +\frac{1}{8(d-3)}\,\xi\right)
         \bigg)\;,\\
   p_{\rm h}^{[m]}&=
      g^4\dA\CA^2\bigg(
         \Iba\frac{d-2}{64}\,\xi^2
         +\Ibb\left(
         -\frac{(d-2)(3d-7)}{768}\,\xi^2
         +\frac{5}{192}\,\xi
         -\frac{1}{32}\right)
-\nonumber\\&
         -\Ibd\frac{d-3}{32}\,\xi^2
         -\Ibe\frac{1}{64}\,\xi^2
         \bigg)\;,\\
\la{eq:n}
   p_{\rm h}^{[n]}&=
      g^4\dA\CA^2\bigg(
         \Iba\!\left(-\frac{d(d\!-\!5)}{32}\,\xi^3
         +\frac{d(d\!-\!10)}{16}\,\xi^2
         +\frac{31\!+\!22 d\!-\!20 d^2\!+\!3 d^3}{4(d-3)(d-5)}\,\xi
         +\frac{2d\!-\!3}{4}\right)
+\nonumber\\&
         +\Ibb\left(\frac{(3d-1)(d-4)}{384}\,\xi^2
         +\frac{22-15 d+3 d^2}{48(d-3)}\,\xi
         +\frac{3+d}{16}\right)
+\nonumber\\&
         +\Ibc\frac{1-2 d}{8}
         +\Ibd\left(\frac{d-3}{8}\,\xi^2
         -\frac{1}{2}\,\xi\right)
         +\Ibe\left(\frac{1}{16}\,\xi^2
         -\frac{1}{4(d-3)}\,\xi\right)
         \bigg)\;,\\
   p_{\rm h}^{[o]}&=
      g^4\dA\CA^2\bigg(
         \Iba\frac{2\!-\!d}{32}\,\xi^2
         -\Ibb\!\left(\frac{1}{96}\,\xi^2
         -\frac{1}{24}\,\xi
         +\frac{1}{8}\right)
         +\Ibd\frac{d\!-\!3}{16}\,\xi^2
         +\Ibe\frac{1}{32}\,\xi^2
         \bigg),\\
   p_{\rm h}^{[p]}&=
      g^4\dA\CA^2\bigg(
         \Iba\left(\frac{1}{16}\,\xi^2
         -\frac{1}{8}\,\xi
         -\frac{1}{8}\right)
         +\Ibb\left(\frac{1}{64}\,\xi^2
         -\frac{1}{32}\,\xi\right)
         +\Ibc\frac{1}{16}
         \bigg)\;,\\
   p_{\rm h}^{[q]}&=
      g^4\dA\CA^2 d\bigg(
         \frac{2+33 d-12 d^2+d^3}{384}\,\xi^4
         -\frac{1+14 d-3 d^2}{32}\,\xi^3
+\nonumber\\&
         +\frac{1+15 d-2 d^2}{16}\,\xi^2
         -d\,\xi
         +\frac{d(d+1)}{4}
         \bigg)\Iba\;,\\
   p_{\rm h}^{[r]}&=
      g^4\dA\CA^2\bigg(
         -\frac{d(2+33 d-12 d^2+d^3)}{192}\,\xi^4
         +\frac{d(7+27 d-6 d^2)}{32}\,\xi^3
-\nonumber\\&
         -\frac{d(11+29 d-4 d^2)}{16}\,\xi^2
         +\frac{2-21 d-33 d^2+8 d^3}{4(d-5)}\,\xi
         -\frac{3d^2}{2}
         \bigg)\Iba\;,\\
   p_{\rm h}^{[s]}&=
      g^4\dA\CA^2\bigg(
         \frac{d(d-5)}{32}\,\xi^3
         -\frac{d(d-9)}{16}\,\xi^2
         +\frac{2+13 d-3 d^2}{4(d-5)}\,\xi
         +\frac{d}{2}
         \bigg)\Iba\;.
\end{align}

\begin{figure}
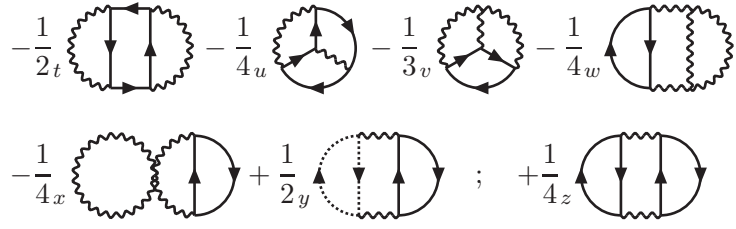

\begin{eqnarray*}
&&
 -{1\over2}_{t}\ToprVBB(\Agl,\Luq,\Agl,\Luq,\Luq,\Luq)
 -{1\over4}_{u}\ToprVM(\Agl,\Auq,\Auq,\Lgl,\Luq,\Luq)
 -{1\over3}_{v}\ToprVM(\Agl,\Auq,\Agl,\Luq,\Luq,\Lgl)
 -{1\over4}_{w}\ToprVBB(\Agl,\Lgl,\Auq,\Lgl,\Lgl,\Luq)
\\[5mm]&&
 -{1\over4}_{x}\ToprVBT(\Auq,\Agl,\Agl,\Agl,\Luq)
 +{1\over2}_{y}\ToprVBB(\Auq,\Lgl,\Ahg,\Lgl,\Luq,\Lhg)
\quad;\quad
 +{1\over4}_{z}\ToprVBB(\Auq,\Lgl,\Auq,\Lgl,\Luq,\Luq)
\end{eqnarray*}
\caption{Three-loop diagrams contributing to $\Nf$, and to $\Nf^2$
(last diagram only).}
\la{fig:3loopNf}
\end{figure}

The three-loop diagrams depicted in \fig\nr{fig:3loopNf} can be expressed 
as (diagram $u\sim\dA/\Nc=\dA(\CA-2\CF)$ 
contributes to two color structures)
\begin{align}
\la{eq:bareNf1}
   p_{\rm h}^{[t+u]}&=
      g^4\dA\Nf\CF\frac{1-d}4\,\bigg(
2(d-1)(\Ifb-2\Ifc+\Ife+\Ifj)
+4\Iff
+(d-5)\Ifg
         \bigg)\;,\\
   p_{\rm h}^{[u+v+w]}&=
      g^4\dA\Nf\CA\bigg(
         \left(\frac{2(d-1)}{d-5}\,\xi
         +(5d-3)\right)\Ifa
         -\frac{2-4 d+d^2}{4}\,\Iff
+\nonumber\\&
         +\frac{(d-1)(d-5)}{8}\,\Ifg
         -\frac{2d-1}{2}\,\Ifh
         \bigg)\;,\\
   p_{\rm h}^{[x]}&=
      g^4\dA\Nf\CA(1-d)\bigg(
         \frac{2}{d-5}\,\xi
         +d
         \bigg)\Ifa\;,\\
   p_{\rm h}^{[y]}&=
      g^4\dA\Nf\CA\bigg(
         -2\Ifa
         -\frac{1}{4}\,\Iff
         +\frac{1}{2}\,\Ifh
         \bigg)\;.
\end{align}

The last three-loop diagram depicted in \fig\nr{fig:3loopNf} gives
\begin{align}
\la{eq:bareNf2}
   p_{\rm h}^{[z]}&=
      g^4\dA\Nf^2\bigg(
         (d-5)\Ifd
         +\frac{d-3}{4}\,\Ifg
         +\Ifi
         \bigg)\;.
\end{align}

Summing all diagrams, 
the gauge parameter $\xi$ explicitly drops out
(and so do the 3-loop sum-integrals $\Ibd$ and $\Ibe$),
leaving the (bare) result
\begin{align}
\la{eq:bareAll}
   p_{\rm h}(T)&=
\frac1d\((1-d)\dA+4\big(2^{-d}-1\big)\Nf\CA\)\Ib{1}{2}
+\nonumber\\&+
      g^2\dA\frac{1-d}{4}\(
          (d-1)\CA
          +2\big(3+4^{2-d}-2^{4-d}\big)\Nf
        \)\Ib{1}{0}\Ib{1}{0}
+\nonumber\\&+
      g^4\dA\CA^2\frac{(1-d)^2}{8}\,\bigg(
          2(d-5)\Iba 
         +\Ibb
         +2\Ibc
         \bigg)
+\nonumber\\&+
      g^4\dA\Nf\CA\frac{1-d}{8}\,\bigg(
         8(d-5)\Ifa
         +2(d-3)\Iff
         -(d-5)\Ifg
         +8\Ifh
         \bigg)
+\nonumber\\&+
      g^4\dA\Nf\CF\frac{1-d}{4}\,\bigg(
         2(d-1)(\Ifb-2\Ifc+\Ife+\Ifj)
         +4\Iff
         +(d-5)\Ifg
         \bigg)
+\nonumber\\&+
      g^4\dA\Nf^2\frac{1}{4}\,\bigg(
         4(d-5)\Ifd
         +(d-3)\Ifg
         +4 \Ifi
         \bigg)
+{\cal O}(g^6)\,.
\end{align}
Note that also the spurious poles (at $d=3-2\e$) which are present
in \eqs\nr{eq:hijk}, \nr{eq:l} and \nr{eq:n} (and which would have severely
limited our ability to evaluate the constant term, requiring 3-loop
sum-integrals to higher order in $\e$) have canceled completely.
Comparing with the corresponding expression given in
\eq(31) of \cite{Braaten:1995jr}, we note complete agreement,
up to two typos in that reference (in their 3-loop terms, 
$\CA T_{\rm F}{\cal I}_1{\cal I}_2\tilde{\cal I}_1$ is
missing a prefactor $(1\!+\!\e)$, 
and $\CF T_{\rm F}{\cal I}_1\tilde{\cal I}_1\tilde{\cal I}_2$
should be multiplied by $1/2$; these correspond to our structures
$\Nf\CA\Ifa$ and $\Nf\CF\Ifc$ above).

Renormalizing the coupling in the $\overline{\rm MS}$ scheme via 
\begin{align}
g^2=\mu^{-2\e}\,g_{\rm bare}^2
=\gR^2\Big(1-\frac{\beta_0}{\e}\,\frac{\gR^2}{16\pi^2}
+{\cal O}(\gR^4)\Big)\;,\quad
\beta_0=\frac{11\CA-2\Nf}{3}
\end{align}
from which (using that $g_{\rm bare}^2$ does not depend on the 
renormalization scale $\mu$) 
its running $\gR^2(\mu)=\gR^2(\mu_0)
\big(1+2\beta_0\frac{\gR^2}{16\pi^2}\ln\frac{\mu_0}\mu+{\cal O}(\gR^4)\big)$ 
follows, and expanding the master sum-integrals around $d=3-2\e$
(see Appendix),
the result coincides
with the expression given in \eq(32) of \cite{Braaten:1995jr}
(note that they write the $\overline{\rm MS}$ scale parameter as $\Lambda$,
where $\Lambda^2\equiv\bar\mu^2\equiv 4\pi e^{-\gaE}\,\mu^2$),
which has subsequently been used to build $p_{\rm QCD}(T)$ 
according to \eq\nr{eq:p_hsu}.
We will not re-iterate discussions about convergence as well
as (renormalization) scale dependence here,
but instead only refer to the 
literature \cite{Arnold:1994ps,Braaten:1995ju,Kajantie:2000iz,Hietanen:2008tv}.
Our main point was the re-derivation of $p_{\rm hard}(T)$ in 
\eq\nr{eq:bareAll}, demonstrating gauge-parameter independence
and using automated computer-algebra methods, which are 
naturally extensible to higher orders.

%
\section{Conclusions}
\la{se:conclusion}

We have re-derived the results of the original 3-loop
computation \cite{Braaten:1995jr} for the QCD pressure.
While the original calculation was performed in Feynman 
gauge, we have used covariant gauge and shown explicit
cancellation of the gauge parameter.
As a slight improvement over \cite{Braaten:1995jr}, our 
IBP reduction revealed that the basis of non-trivial
3-loop master integrals used
in the original computation could be reduced by one. 

In future work, we hope to be able to employ the IBP setup
to the 4-loop level, thus contributing to the last missing 
piece needed for the {\em physical leading-order} (i.e. $g^6$) determination
of the pressure within the dimensionally reduced effective theory framework
\cite{Braaten:1995cm,Kajantie:2000iz,Hietanen:2004ew} 
-- the only currently known framework allowing
for a weak-coupling expansion of thermodynamic observables
that is systematically improvable.

%
\acknowledgments

We are indebted to Shin-ichiro Hara for advice on his algebraic 
library for Ruby.
M.N.~thanks the theory group of Bielefeld University for 
hospitality during the period when this work was done.
The work of Y.S.~is supported by the Heisenberg Programme of the Deutsche
Forschungsgemeinschaft (DFG grant SCHR 993/1-1).
All figures were drawn with Axodraw~\cite{Vermaseren:1994je}.

\appendix

%
\section{One- and two-loop vacuum sum-integrals}
\la{app:12loop}

The one-loop bosonic tadpole is known analytically and reads
($d=3-2\e$)
\begin{align}
\la{eq:Imn}
\Ib{m}{n} \equiv \sumint{P}\frac{P_{0}^{n}}{(P^{2})^{m}} 
 = \frac{2 \pi^{3/2} T^4}{(2\pi T)^{2m-n}}\(\frac{\mu^2}{\pi T^2}\)^{\e}
 \frac{\Gamma\(m - \frac{3}{2} + \e\)}{\Gamma(m)}\,\zeta(2m - n - 3 + 2\e)\,,
\end{align}
whereas the fermionic tadpole can be related to the corresponding bosonic 
one via
\begin{align}
\la{eq:Ifmn} 
\If{m}{n} \equiv \sumint{\{P\}}\frac{P_{0}^{n}}{(P^{2})^{m}} 
 = (2^{2m-n-d} - 1)\Ib{m}{n}\,,
\end{align}
which immediately follows from the scaling relations explained 
in \se\ref{se:reduction}.

As mentioned above, via integration-by-parts relations
all two-loop integrals are expressible in terms of 
products of two one-loop tadpoles which means they are also available
analytically up to arbitrary order in $\e$.
A typical reduction, following from IBP and being used in the 
2-loop contribution to $p_{\rm hard}(T)$, reads
\begin{align}
\la{eq:S0}
{\cal S}&\equiv\sumint{PQ}\frac{1}{P^2\,Q^2\,(P-Q)^2}\;=\;0\;.
\end{align}
This remarkable result in fact follows from the IBP relation
\begin{align}
&0=\sumint{PQ}\partial_{p_i}\,f_i\,\frac{1}{P^2\,Q^2\,(P-Q)^2}\\
&\mbox{with~~}f_i=(d-2)(p_i+q_i)+\frac{2}{Q^2}\,(P_0+Q_0)(q_i P_0 -p_i Q_0)\;,
\end{align}
from which, after working out the derivatives, 
using the shift $Q\rightarrow P-Q$ 
and exploiting the $P\leftrightarrow Q$ symmetry, it follows that
$0=(d-2)(d-3){\cal S}$, proving \eq\nr{eq:S0}.
Note that in the literature, by explicit integration
\eq\nr{eq:S0} was only known to
hold to ${\cal O}(\e)$ \cite{Arnold:1994ps}.

%
\section{Three-loop vacuum sum-integrals}
\la{app:3loop}

All non-trivial three-loop master sum-integrals that are needed for
$p_{\rm hard}$ of \eq\nr{eq:bareAll} have been evaluated
in Ref.~\cite{Arnold:1994ps}, and been subsequently summarized in
the literature \cite{Braaten:1995jr,Schroder:2008ex}.
In the notation of \se\ref{se:result}, they read
\begin{align}
\Ibb &= \frac{T^4}{16\pi^2}\(\frac{\mu^2}{4\pi T^2}\)^{3\e}\frac1\e\,
\frac{1}{24}\Big[ 1+\(\fra{91}{15}-3\gaE+8\zA-2\zC\)\e+{\cal O}(\e)^2\Big]\;,\\
\Ibc &= \frac{T^4}{16\pi^2}\(\frac{\mu^2}{4\pi T^2}\)^{3\e}\frac1\e\,
\frac{11}{216}\Big[ 1+\(\fra{73}{22}-\fra{21}{11}\gaE+\fra{64}{11}\zA-\fra{10}{11}\zC\)\e+{\cal O}(\e)^2\Big]\;,\\
\Ifg &= \frac{T^4}{16\pi^2}\(\frac{\mu^2}{4\pi T^2}\)^{3\e}\frac1\e\,
\frac{1}{96}\Big[ 1+\(\fra{173}{30}-3\gaE-\fra{42}5\ln2+8\zA-2\zC\)\e+{\cal O}(\e)^2\Big]\;,\\
\Ifh &= \frac{T^4}{16\pi^2}\(\frac{\mu^2}{4\pi T^2}\)^{3\e}\frac1\e\,
\frac{-29}{1728}\Big[ 1+\(\fra{89}{29}-\fra{39}{29}\gaE-\fra{90}{29}\ln2
+\fra{136}{29}\zA-\fra{10}{29}\zC\)\e+{\cal O}(\e)^2\Big]\;,\\
\Ifi &= \frac{T^4}{16\pi^2}\(\frac{\mu^2}{4\pi T^2}\)^{3\e}\frac1\e\,
\frac{1}{108}\Big[ 1+\(\fra{35}{8}-\fra32\gaE-\fra{63}{10}\ln2+5\zA-\fra12\zC\)\e+{\cal O}(\e)^2\Big]\;,\\
\Ifj &= \frac{T^4}{16\pi^2}\(\frac{\mu^2}{4\pi T^2}\)^{3\e}\frac1\e\,
\frac{-1}{192}\Big[ 1+\(\fra{361}{60}+3\gaE+\fra{76}5\ln2-4\zA+4\zC\)\e+{\cal O}(\e)^2\Big]\;,
\end{align}
where $Z_i\equiv\frac{\zeta'(-i)}{\zeta(-i)}$.
$\Iff$ can be obtained via \eq\nr{eq:PCrel}, and coincides
with the expansion given in \cite{Braaten:1995jr}.

%

\end{document}